\documentclass[
 reprint,
 amsmath,amssymb,
 aps,
pra,
]{revtex4-2}

\usepackage{graphicx}% Include figure files
\usepackage{dcolumn}% Align table columns on decimal point
\usepackage{graphicx}%
\usepackage{multirow}%
\usepackage[inline]{enumitem}
\usepackage{xcolor}%
\usepackage{physics}
\usepackage{comment}
\usepackage{ulem}

\newtheorem{lemma}{Lemma}

\usepackage{hyperref}
\usepackage{pifont}
\newcommand{\fpo}{\mathcal{L}^{\dagger}}
\newcommand{\fpoa}{\mathcal{L}}
\newcommand{\D}{\mathcal{D}(t)}
\newcommand{\DT}[1]{\mathcal{D}_{#1}(t)}
\newcommand{\DTo}[1]{\mathcal{D}_{#1}(0)}
\newcommand{\req}{\rho_{eq}(x)}
\newcommand{\rxt}{\rho(x,t)}
\newcommand{\lpi}{L^2_{\pi}(\Omega)}

\providecommand{\dd}{\mathrm{d}}

\newcommand{\pcmd}[1]{\textcolor{red}{#1}}
\newcommand{\iitmd}[1]{\textcolor{purple}{#1}}

\begin{document}
\preprint{APS/123-QED}

\title{Mathematical analysis of the Markovian Mpemba effect}% 

\author{Umashankar Pardhi$^{1}$}
%\thanks{d23020@students.iitmandi.ac.in}
\author{Syed Abbas$^{2}$}
%\thanks{abbas@iitmandi.ac.in}
%\author{Harsh Soni$^{1}$}
%\email{harsh@iitmandi.ac.in}
\author{Moupriya Das$^{3}$}
\email{moupriya@iitmandi.ac.in}

\affiliation{
	$^1$School of Physical Sciences, Indian Institute of Technology Mandi, Mandi 175001, India \\
    $^2$School of Mathematical and Statistical Sciences, Indian Institute of Technology Mandi, Mandi 175001, India \\
    $^3$School of Chemical Sciences, Indian Institute of Technology Mandi, Mandi 175001, India\\
	}
\date{\today}
%\pacs{05.40.-a, 05.70.Ln,  45.70.Vn}

\begin{abstract}
The Mpemba effect is a counterintuitive phenomenon in which two equilibrium systems at different temperatures, say hot and warm systems, when quenched at the same bath temperature, for some pair of hot and warm temperatures, the hot system comes to bath temperature faster than the warm system. In recent years, studies on the Mpemba effect for Markovian systems under external potential, which are continuous-time, continuous-state-space Markov processes, have been carried out theoretically and experimentally by resorting to pseudo-distance measures which are not true distance metrics. Moreover, maximum crossover time is given for the Mpemba effect for all monotone distance measures or functions (which include true distance metrics and pseudo-distance measures). However, no mathematical analysis is carried out for the mere possibility of the presence of the Mpemba effect for Markovian systems under an external potential with a true distance metric. In this paper, by using a true distance metric which arises naturally from the weighted Hilbert space in which the adjoint of the generator of the Markov process is a Hermitian operator, we have shown the mathematical analysis of the possibility of the presence of the Mpemba effect by using Poincar\'e inequality and the geometry of Hilbert space.\\
\end{abstract}

%\keywords{Suggested keywords}%Use showkeys class option if keyword
                              %display desired
\maketitle

%\tableofcontents
%%%%%%%%%%########
%%%%%%%%%%%%######
\begin{comment}
    A function that satisfies the triangle inequality guarantees that the direct path between two points is never longer than a detour through a third point. This key property ensures a stable, predictable geometric space where shortest paths exist, distances are consistent, and algorithms like clustering or nearest-neighbor searches work reliably.
    In mathematics, majorization is an ordering relation that compares the spread, or inequality, of the components of two real vectors. If vector y is majorized by vector x (written as \(y \prec x\)), it means the components of y are more evenly distributed, or less spread out, than the components of x, while both vectors share the same sum.
\end{comment}
\section{Introduction}
%\mdas{It is a general perception that a colder object will require less time to cool down to the same lower temperature compared to a hotter one when all other conditions remain the same. However, an interesting counterintuitive phenomenon in the case of the cooling of water is observed in which hot water cools faster than warm water. Historically, the first observation of this phenomenon was reported by Aristotle in 350 BCE and was systematically shown by a series of experiments by Mpemba and Osborne in the 1960s for water \pcmd{[Ref: \textit{Nature Reviews Physics} \textbf{3}, 534–535 (2021) ]}. It was found later that this occurrence, which is now termed the Mpemba effect, is not restricted to water only. In contrast, it emerges in a wide variety of systems \pcmd{[Refs.]}. The analysis of this effect is important both for fundamental understanding and application purposes involving this anomalous relaxation process in which the intermediate states remain far away from equilibrium during the quenching procedure. However, proposing a generalized underlying mechanism for this effect still exists as an open question \pcmd{[Ref. \textit{Nature} \textbf{584}, 64–68 (2020)]}.}
It is a general perception that a colder object will require less time to cool down to the same lower temperature compared to a hotter one when all other conditions remain the same. However, an interesting counterintuitive phenomenon in the case of the cooling of water is observed in which hot water cools faster than warm water. Historically, the first observation of this phenomenon was reported by Aristotle in 350 BCE and was systematically shown by a series of experiments by Mpemba and Osborne in the 1960s for water \cite{john}. It was found later that this occurrence, which is now termed the Mpemba effect, is not restricted to water only. In contrast, it emerges in a wide variety of systems \cite{teza, naturereview}. The analysis of this effect is important both for fundamental understanding and application purposes involving this anomalous relaxation process in which the intermediate states remain far away from equilibrium during the quenching procedure. However, proposing a generalized underlying mechanism for this effect still exists as an open question \cite{avinash}.

There have been significant advances in the development of the microscopic theory of the Mpemba effect, which is effectively realized in numerous macroscopic systems \cite{mazo}.% \pcmd{[Ref. Phys. Rev. Lett. 134, 107101 (2025)]}. 
The idea is that the interpretation of the detailed mechanism of this completely nonequilibrium relaxation process can be possible within the purview of the microscopic framework \cite{mindex}. While many specific microscopic theories and methodologies have been proposed in the present context, a general theoretical scheme to explain the basics of this phenomenon is missing.

The major microscopic theories on the classical Mpemba effect are centred on Markovian systems, which provide the minimal setup for nonequilibrium systems. The usual technique, which has been adopted in most of the studies, is to evaluate the deviation of the system from its equilibrium or stationary state during the time of the relaxation through a quantity termed a distance measure. A large variety of possible distance measures has been introduced to explain the Mpemba effect in different contexts~ \cite{luraz, avinash,apurbasingle,isha}. While these quantities satisfy the monotonicity and convexity conditions \cite{luraz}% \pcmd{[Ref. PNAS, 114 (20) 5083-5088 (2017)]}
, they do not fulfil the triangular inequality criterion. Therefore, they are considered as `pseudo-distance measures', not as the `true distance metrics'. In different situations, distinct distance measures are used to interpret the Mpemba effect, identifying the advantage of one over the other. However, there is no rigorous logical understanding of the selection of a particular measure. Moreover, if the existence of the Mpemba effect is aimed to be investigated from a general point of view, all monotone measures are required to be analyzed within a unified theoretical framework. This universal approach has remained unaddressed from the perspective of distance measures.

For a continuous state-space Markovian system using a pseudo-distance measure, different combinations of conditions on an external one-dimensional potential for the presence or absence of the Mpemba effect, such as metastability, domain asymmetry, hard wall (i.e., a bounded domain), and no wall (i.e., an unbounded domain), are given \cite{tanwall,predictingconditions,avinash, apurbadouble, apurbasingle}.
Also, for a two-dimensional bistable potential, the presence of the Mpemba effect is reported \cite{twodimensional}. For the case of a discrete state-space Markovian system, different conditions for the presence of the Mpemba effect are discussed in \cite{mindex, necessaryconditions}. For both types of Markovian systems, i.e., Markov chain and continuous state-space case, Vu et al.\cite{mazo} have introduced the concept of thermomajorization to analyze the Mpemba effect. In this study, they have reported an upper bound of crossover time, i.e., the time after which a system initially with a higher temperature will certainly relax faster than a relatively warm system for all monotone pseudo-distance measures and true distance metrics. 
They have calculated the crossover time for a three-state Markovian system but not for a continuous state-space system. 
%\iitmd{[MD: Note: Markovian systems are considered as exact calculations of the distance measures are possible for these systems. For classical systems, the works are mainly based on Markovian systems. Interpretation of the quantum Mpemba effect has more work for non-Markovian systems. Noted for Reviewer's response if required in the future.]}

Though great progress has been made for the quantification of the Mpemba effect in Markovian systems, in the first place, no analysis is available for the mere possibility of the Mpemba effect in Markovian systems by mathematical analysis treatment. We will show, by functional analysis, the possibility of the Mpemba effect in Markovian systems. We have used the Poincar\'e inequality of the Gibbs measure and the geometry of Hilbert space to show the possibility of the Mpemba effect in Markovian systems.

In this work, we consider a system connected to a heat bath such that the system's dynamics is given by a continuous-time, continuous-state-space Markov stochastic process which can be represented by an overdamped Langevin equation.
%\mdas{The dynamics is also governed by an external potential.} 
Corresponding to this stochastic process, we get weighted Hilbert spaces in which the Fokker-Planck operator corresponding to an overdamped Langevin equation is Hermitian. In this Hilbert space, we naturally get a true distance metric which satisfies the triangular inequality as well. This is induced from the inner product structure of Hilbert spaces. Therefore, our study satisfies the generality aspect of the analyses in terms of all monotone measures through the estimation of the true distance metrics. This framework has quantitative rigidity over the thermomajorization technique, which is primarily based on the ordering relation. Consequently, our technique can be considered to provide a more quantitative and descriptive outlook in studying the dynamical aspects of the Mpemba effect. Finally, the present work incorporates the idea of the physical potential in continuous space in the analyses of the Mpemba effect. Thus, it bears the advantage of interpreting the easier connection of the phenomenon to real systems and experiments, including barrier-crossing and several other relevant nonequilibrium processes, and helps understand the important control over the effect by the underlying potential. %making it feasible to introduce the effect of the external force directly, and also leaving the scope for stochastic thermodynamic calculations. (These two aspects may be mentioned in the Conclusion.)
The last leverage has already been understood as the benefit of considering continuous state-space  Markovian systems with external potential over Markov chains in connection to the Mpemba effect from the earlier studies.

The manuscript is organized as follows. In section: \ref{sec:setup}, we discuss Markovian process of our interest and Hilbert spaces in which the concerned operators are Hermitian. In section: \ref{sec:metric}, we describe distance metrics and their properties. In section: \ref{sec:result}, we depict the results. We conclude the manuscript in section: \ref{sec:conclusion}.
\begin{comment}
\iitmd{[MD: Structure of the Introduction: Definition (of Mpemba effect), Discussion of the existing studies and analyses, The gap, Introduction of the discussion about the present study and how it fills the gap, Road-map of the manuscript]}
\pcmd{[MD: General suggestion is to use the terms `pseudo-distance measure' and `true distance metric' for distinction]}
\end{comment}
%%%%%%%%%%%
%%%%%%%%%%%%%%
%%%%%%%%%%%%%%
\section{Setup and theory}\label{sec:setup}
In this section, we discuss Markovian processes which follow the overdamped Langevin equation and Hilbert spaces in which operators such as the generator of the process and the Fokker-Planck operator, which is the adjoint of the generator, are Hermitian.

%%%%%%%%%%%%%======
%\subsection{Fokker-Planck Equation corresponding to overdamped Langevin equation}\label{sec:FokkerPlanck}
The overdamped Langevin equation for diffusion processes under an external potential is given by
\begin{equation}\label{langevin}
    \frac{d\tilde{x}}{d\tilde{t}} = -\frac{D}{k_B T_b}\frac{d\tilde{V}(\tilde{x})}{d\tilde{x}} + \sqrt{2D}\xi(\tilde{t}),
\end{equation}
where \(D\) is the diffusion constant, \(k_B\) is the Boltzmann constant, \(T_b\) is the bath temperature, the external potential \(\tilde{V}(\tilde{x}) = k_B T_b \, V(\tilde{x}/l)\) , \(l\) is constant 
%\pcmd{[MD: To specify the constant. It is better to connect it with a characteristic length of the system, if possible.]}
having the dimension of length, and \(\xi(t)\) is Gaussian white noise  
%\pcmd{[MD: To refer to the zero-mean property of this noise before presenting the correlation property.]}
with \(\langle \xi(\tilde{t}) \rangle = 0\)  and \(\langle \xi(\tilde{t}), \xi(\tilde{t}') \rangle = \delta(\tilde{t}-\tilde{t}')\). The domain of potential, which is also domain of \(\tilde{x}\), can be bounded with a hard wall (i.e., \(\tilde{x} \in (a,b) \subset \mathbb{R}\), and a hard wall at \(\tilde{x} = a\) and \(\tilde{x} = b\)), or unbounded i.e., \(\tilde{x} \in \mathbb{R}\). In case of unbounded domain, potential is confining i.e. \(\lim_{\tilde{x} \to \pm\infty} (\tilde{V}'(\tilde{x}))^2/2 - \tilde{V}''(\tilde{x}) = \infty \). These conditions on the potential are called reflecting boundary conditions. For non-dimensional variable \(x = \tilde{x}/l\) and \(t = \tilde{t} D/l^2\) the overdamped Langevin equation \eqref{langevin} reduces to a non-dimensionalised Langevin equation
\begin{equation}\label{langevinnondim}
    \frac{dx}{dt} = -\frac{dV(x)}{dx} + \sqrt{2}\eta(t),
\end{equation}
where \(\eta(t)\) is the Gaussian white noise  with \(\langle\eta(t') \rangle = 0\) and \(\langle\eta(t'), \eta(t) \rangle = \delta(t-t').\)
Corresponding to this overdamped non-dimensionalized Langevin equation (\ref{langevinnondim}), we have the Fokker-Planck equation for the evolution of probability density \(\rho(x,t)\) \cite{Pavliotischap2, Holmeschap10, lichap8,Balakrishnan2021} 
\begin{equation}\label{FPE}
\frac{\partial }{\partial t} \rho(x,t) = \frac{\partial}{\partial x} \Bigl(\frac{dV(x)}{dx} \rho(x,t) \Bigr) + \frac{\partial^2}{\partial x^2} \rho(x,t).
\end{equation}
In operator form, the Fokker-Planck equation \eqref{FPE} can be written as
\begin{equation}\label{FPEonOPERATORform}
\frac{\partial }{\partial t} \rho(x,t) =\fpo \rho(x,t),
\end{equation}
where \(\fpo\) is the Fokker-Planck operator, expressed as
\begin{equation}\label{FPoperator}
\fpo = \frac{\partial}{\partial x} \frac{dV(x)}{dx} + \frac{\partial^2}{\partial x^2}.
\end{equation}
\begin{comment}

The continuity form of the Fokker-Planck equation, which is a conservation equation of probability density,  \(\rho(x,t)\) is given as
\begin{equation}\label{FPEcontinuityForm}
    \frac{\partial }{\partial t} \rho(x,t)+ \frac{\partial}{\partial x} j(x,t) = 0,
\end{equation}
where \(j(x,t)\) is called the 'probability flux' or 'probability current' and is \(\fpo\), which tells us about how probability moves, and is given by
\begin{equation}\label{current}
    j(x,t) = - \frac{dV(x)}{dx} \rho(x,t) - \frac{\partial}{\partial x} \rho(x,t) .
\end{equation}

For reflecting boundary conditions, in the case of a bounded domain there will be no probability current across the boundary, and for an unbounded domain, \(\lim_{x \to \pm\infty} j(x,t) = 0\) \cite{Pavliotischap4, Holmeschap10}. This leads to the unique stationary distribution \cite{Holmeschap12, lichap8}, also called a Boltzmann distribution or Gibbs measure, given by
\end{comment}
For reflecting boundary conditions \cite{Pavliotischap4, Holmeschap10}, we have the unique stationary distribution \cite{Holmeschap12, lichap8}, also called a Boltzmann distribution or Gibbs measure, given by
\begin{equation}
    \rho_{eq}(x) = Z^{-1}  e^{\frac{-V(x)}{k_B T}},
\end{equation}
where \(Z= \int_{\Omega} \exp[{-V(x)/(k_B T)}]dx\) and \(\Omega\) is a domain. 
Corresponding to Eq.~\eqref{langevinnondim}, in addition to the Fokker-Planck equation  Eq.~\eqref{FPE}, we have the Kolmogorov backward equation given by
\begin{align}\label{KBE}
    \frac{\partial g(x,t)}{\partial t} &= -\frac{dV(x)}{dx} \frac{\partial g(x,t)}{\partial x} + \frac{\partial^2 g(x,t)}{\partial x^2} \\
    &= \fpoa \, g(x,t), \nonumber
\end{align}
where \(g(x,t) = \rho(x,t)/\rho_{eq}(x) \). \(\fpoa\) is called the generator of diffusion process \eqref{langevinnondim}.
%If vector space \(W\) is a weighted complete inner product (i.e. Hilbert space) with a weight \(s(x)\) and  inner product \(\braket{f}{h}_{s} = \int_{\Omega} \, f(x)\,h(x) \,s(x) dx\), the weight \(s(x)\) is a positive function and \(\int_{\Omega} \, s(x) < \infty\), then we denote Hilbert space \(W\) as \(L^2_s(\Omega)\).
For reflecting boundary conditions,  we have the following results \cite{Holmeschap12, Pavliotischap4}:
%\[j(x,t)= 0 \iff \fpoa \, \text{is self adjoint in}\,  L^2_{\pi}(\Omega) \iff X \, \text{satisfies detailed balance}\]

%\[j(x,t)= 0 \iff \fpo \, \text{is self adjoint in}\,  L^2_{\omega}(\Omega) \iff X \, \text{satisfies detailed balance}\]

%\[j(x,t)= 0 \implies \fpo (g(x) \rho_{eq}(x) = \rho_{eq}(x) \fpoa g(x)\]
\begin{itemize}
    \item \(\fpoa\) is Hermitian and negative semi-definite in Hilbert space \(L^2_{\pi}(\Omega)\)
    \item \(\fpo\) is Hermitian and negative semi-definite in Hilbert space \(L^2_{\omega}(\Omega)\)
    \item \(\fpo (g(x) \rho_{eq}(x)) = \rho_{eq}(x) \fpoa g(x)\),
    %\item \(X\) satisfies detailed balance,
\end{itemize}
where %\(X\) is the stochastic process of equation(\ref{langevinnondim}),
\(\pi(x)= \rho_{eq}(x)\),  \(\omega(x) = 1/\rho_{eq}(x)\).%, and \(g(x) \in C^2(\Omega)\)

%\begin{comment}
 Since both operators \(\fpo\), \(\fpoa\) are self-adjoint (or Hermitian) in their respective Hilbert spaces, both have real eigenvalues and  orthonormal eigenvectors.
Both operators have the same eigenvalues, and all eigenvalues are non-positive because both operators are negative semi-definite. 
Say, \(\{\lambda_i\}_{i=0}^{\infty}\) for eigenvalues of \(\fpo\) and \(\fpoa\), \(\{\phi_i\}_{i=0}^{\infty}\) for orthonormal eigenfunctions of \(\fpoa\), and \(\{\psi_i\}_{i=0}^{\infty}\) for orthonormal eigenfunctions of \(\fpo\). 
Eigenfunctions follow relation \(\psi_i(x)= \rho_{eq} \, \phi_i(x) \), and \(\phi_i, \psi_i\) form a bi-orthogonal set in \(L^2\) (the non-weighted inner product space), i.e., \(\braket{\phi_i}{\psi_i}=\delta_{ij}\), \(\delta_{ij}\) is the Kronecker delta.
Eigenvalues are ordered as \(0=\lambda_1 > \lambda_2 \geq \lambda_3 \geq \lambda_4 \ldots\), and satisfy \(\lambda_i \to -\infty\) as \(i \to \infty\). 
The largest eigenvalue \(\lambda_1=0\) has multiplicity, \(1\) and the corresponding eigenfunctions are \(\phi_1 = 1, \, \psi_1 = \rho_{eq} \)  \cite{Holmeschap12}. 
Then the solution of equation (\ref{FPEonOPERATORform}) will be given by
\begin{align}\label{solution}
    \rho(x,t) &= \sum_{i=1}^{\infty} a_i e^{\lambda_i t} \psi_i(x) \nonumber \\
    &= \rho_{eq}(x) + \sum_{i=2}^{\infty} a_i e^{\lambda_i t} \psi_i(x),
\end{align}
where
\begin{align}
    a_i &= \braket{\rho(x,0)}{\psi_i(x)}_{\omega} \label{coefficientsw} \\
    &= \braket{g(x,0)}{\phi_i(x)}_{\pi} \label{coefficientspi}
\end{align}
\(\rho(x,0)\) is the initial condition for equation (\ref{FPEonOPERATORform}) and \(g(x,0) = \rho(x,0)/\rho_{eq}(x)\). Physically \(a_i\) represents the magnitude of overlap of \(\psi_i(x)\) with the initial condition \(\rho(x,0)\).
   
%\end{comment}

%%%%%%%%%%%%%%%%%%=====

%%%%%%%%%%%%%%
%%%%%%%%%%%%%
\section{Distance metric and its properties}\label{sec:metric}
Here, we discuss the distance metric which is employed to quantify the Mpemba effect in our work.

%%%%%%%%%%%%%%%%%%%%%%%%%%%%%%%%
\subsection{Distance metric corresponding to overdamped Langevin system}\label{subsec:contidistance}
We know from functional analysis that the norm on inner product spaces \((V)\) induced by the inner product and \(\forall h \in V\), is defined as \(\norm{h}= \sqrt{\braket{h}{h}}\). The metric (or distance) induced by the norm is defined as \(\forall f, g \in V\), \(d(f,g) =\norm{f-g}\) \cite{Clason2020}.
So we can define the distance of an instantaneous probability density function from the equilibrium probability density function as 
 \begin{align}\label{eq:metric} %, \nonumber \\
     \D & = d(\rxt, \req)\nonumber \\ 
     &= \norm{\rxt -\req}_{\omega}  \nonumber \\
       &= \left(\int_{\Omega} (\rxt-\req)^2 \, \frac{1}{\req} dx\right)^{1/2}.
 \end{align}
 This distance \(\D\) (or \(d(\rxt, \req)\) ) is induced from the norm of the Hilbert space in which the Fokker-Planck operator \(\fpo\) is self-adjoint (or Hermitian). Distance \(\D\) has the following properties:
 \begin{enumerate}
     \item \label{prop:conticonvex} It is a continuous, non-negative, convex function of \(\rxt\).
     \item \label{prop:zero} \(\D =0\) if and only if \(\rxt = \req\).
     \item \label{prop:triangle} If \(\DT{1} = d(\rho_1(x,t),\req) \), \(\DT{2} = d(\rho_2(x,t),\req) \), and \(\DT{3} = d(\rho_2(x,t),\rho_1(x,t)) \) then \(\DT{3} \leq \DT{1} + \DT{2}\).
     \item \label{prop:monotone} It is monotonically non-increasing with time \(t\).
     \item \label{prop:bigger} If \(T_h > T_w\) then \(\DTo{h} > \DTo{w}\), where \(\DTo{h} \) and \(\DTo{w} \) are distance from the equilibrium of the system with initial temperature \(T_h\) and \(T_w\), respectively.
 \end{enumerate}
 Properties \ref{prop:conticonvex}, \ref{prop:zero}, \ref{prop:triangle} are due to induced distance (metric) from norm \cite{kumaresan2013topology}. Proof of properties \ref{prop:monotone} and \ref{prop:bigger} is given in appendix~\ref{app:oodist}. 

 By considering equations~\eqref{solution} and~\eqref{eq:metric}, we can write distance from equilibrium as
 \begin{equation*}
     \D = \left(\sum_{i=2}^{\infty} a_i^2 e^{2\lambda_i t} \right)^{1/2}.
 \end{equation*}
 In the long time, the distance from equilibrium will be \(\D \approx |a_2| \exp(\lambda_2 t)\). This dictates that \(|\lambda_2|\) is the slowest relaxation rate and is also called the spectral gap, \(\exp(\!-|\lambda_2|\,t)\) the long-time decay factor, and \(\psi_2(x)\) the spatial profile of the slowest mode.

%%%%%%%%%%%%%%%
%%%%%%%%
\subsection{Physical interpretation of the distance metric in a double-well potential}

\begin{figure}%[t]
  \centering
  \includegraphics[width=\columnwidth]{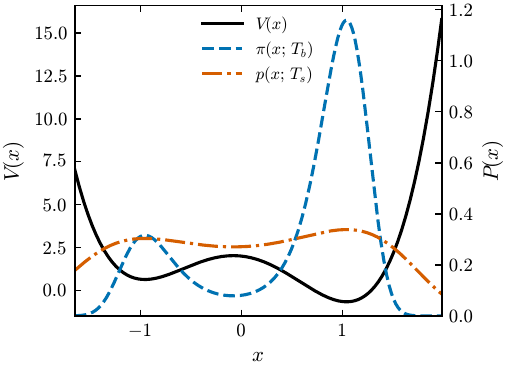}
  \caption{Asymmetric double-well potential
  %$V(x) = 2(1 - x^{2})^{2} - 0.65\,x$ (solid line, left axis)
  , together with the
  equilibrium Gibbs distribution $\rho_{\mathrm{eq}}(x) = Z^{-1} \exp[-V(x)]$ at the
  bath temperature $T_b$ (dashed line, right axis) and the initial distribution
  $\rho(x,0)$ of the system prepared at a higher temperature 
  %$T_s = 12\,T_b$
  $T_s$
  (dash-dotted line, right axis). The equilibrium distribution is concentrated in
  the two wells. The weight $1/\rho_{\mathrm{eq}}(x) = Z \exp[V(x)]$ in the metric
  $\mathcal{D}(t)$ is therefore largest away from the well minima, near the barrier
  and the outer walls, where $\rho(x,0)$ deviates most from $\rho_{\mathrm{eq}}(x)$.
  Energies are measured in units of $k_B T_b$.}
  \label{fig:potpdf}
\end{figure}

The distance metric introduced in
Eq.~\eqref{eq:metric} can be written as
\begin{comment}
\begin{align}
    \mathcal{D}(t)
    &= d(\rho(x,t),\rho_{\mathrm{eq}}(x)), \nonumber\\
    &= \left(\int_{\Omega}
       \frac{\bigl(\rho(x,t)-\rho_{\mathrm{eq}}(x)\bigr)^2}
            {\rho_{\mathrm{eq}}(x)} \,dx \right)^{1/2} \nonumber\\
    &= \left(Z\int_{\Omega}
       \bigl(\rho(x,t)-\rho_{\mathrm{eq}}(x)\bigr)^2
       e^{V(x)} \,dx\right)^{1/2},
\end{align}
where
\(
\rho_{\mathrm{eq}}(x)=Z^{-1}e^{-V(x)}
\)
is the Gibbs equilibrium distribution.
\end{comment}
\begin{align*}
    \mathcal{D}(t) =
    \left(Z\int_{\Omega}
       \bigl(\rho(x,t)-\rho_{\mathrm{eq}}(x)\bigr)^2
       e^{V(x)} \,dx\right)^{1/2},
\end{align*}
where
\(
Z = \int_{\Omega} \exp[{-V(x)}]. \, 
\)
%\pcmd{[MD: The first two equalities have already been described in Eq.~\eqref{eq:metric}. They can be removed. The final equation can be written mentioning that it follows from Eq.~\eqref{eq:metric} as it has been stated in the above sentence.]} 
To understand the physical significance of this metric, we consider a
double-well potential $V(x)$, shown schematically in
Figure~\ref{fig:potpdf}. The potential consists of two wells
separated by an energy barrier. After the system is coupled to a thermal
bath, relaxation toward equilibrium typically occurs on two distinct
timescales. On relatively short timescales, probability redistributes
rapidly within each well (intra-basin relaxation). On much longer
timescales,% \textit{probability} \iitmd{[To replace with]} \mdas{the distribution} 
the distribution crosses the energy barrier, 
allowing the two basins to exchange probability and ultimately reach global equilibrium
(inter-basin relaxation)\cite{luraz}.
The weight appearing in the metric,
\[
\frac{1}{\rho_{\mathrm{eq}}(x)}
    =Ze^{V(x)},
\]
is in the maximum range in regions where the equilibrium probability is negligible.
%\pcmd{[MD: Comment: As ``largest'' or ``smallest'' refer to unique values, they are connected to a point. Here, a range is indicated in terms of the weight factor and the probability, as ``regions'' is mentioned. So, the sentence may be re-framed as, ``.... is in the maximum range in regions where the equilibrium probability is negligible.'']}
For a double-well potential, 
%these are precisely the barrier regions between the metastable wells.
these are precisely the regions away from well minima. 
Consequently, discrepancies between 
$\rho(x,t)$ and $\rho_{\mathrm{eq}}(x)$ %near the barrier contribute
away from well minima contribute more strongly to the distance.
% than comparable discrepancies inside thewells. 
Since barrier crossing governs the slowest relaxation mode of 
the system, the metric naturally emphasizes the physically relevant 
part of the dynamics that controls convergence to equilibrium. 

This interpretation is consistent with the spectral description of the 
Fokker--Planck operator, where the smallest nonzero eigenvalues are
associated with inter-basin transitions. Thus, the weighted Hilbert
space metric is not merely mathematically natural, but also reflects
the dominant physical mechanism underlying long-time relaxation in
multistable systems. In the schematic, an asymmetric double-well potential is shown, though the above facts are true for a symmetric double-well potential as well.

%\iitmd{[MD: It has not been specified whether the double-well potential is asymmetric or symmetric. From the schematic, it appears to be asymmetric. For this framework, is asymmetry in the potential essential to consider? Or does the mathematical treatment also comprise the symmetric case? It is suggested to mention these facts explicitly in the discussion of this subsection.]}

%%%%%%%%%%%
%%%%%%%%
\subsection{Thermomajorization and true distance metric}

%\mdas{In this subsection, we describe in detail the advantage of our true distance metric framework over the thermomajorization technique, which has been introduced recently~\cite{mazo} to interpret the Mpemba effect within the purview of a general analysis.} 
%\iitmd{[MD: Added the above introductory sentence to provide the readers an idea why these two approaches are being compared.]}
In this subsection, we describe in detail the advantage of our true distance metric framework over the thermomajorization technique, which has been introduced recently~\cite{mazo} to interpret the Mpemba effect within the purview of a general analysis.

In mathematics, an ordering relation such as majorization compares vectors by the spread: if $\mathbf{y} \prec \mathbf{x}$, the components of $\mathbf{y}$ are more uniformly distributed than those of $\mathbf{x}$. Majorization is a powerful tool precisely because it encodes a partial order, a structural hierarchy among distributions, without reference to any notion of distance. Thermomajorization extends this idea to thermal systems by incorporating the energy structure of the bath through the Gibbs distribution $\pi$, producing the ordering $p' \prec_\pi p$ that governs which relaxation trajectories are thermodynamically accessible. When the thermomajorization Mpemba effect holds, the hot system's distribution lies below the warm system's Lorenz curve for all time beyond the crossover $t_*$, certifying that the hot state is simultaneously closer to equilibrium than the warm state under every monotone measure at once.

Powerful as this ordering is, it carries an inherent limitation: it tells us which of two distributions is closer to equilibrium, but not to what extent, nor does it provide a number that can be tracked, plotted, or compared across different systems. Just as a ranking of athletes tells you who is faster but not by what margin, thermomajorization orders distributions without furnishing a distance. The Mpemba effect, identified through thermomajorization, thus lacks a natural geometric measure%language \iitmd{[MD: The word ``language'' may be replaced by ``measure''.]}. 
. Mpemba effect is fundamentally a statement about which system reaches a small neighborhood of equilibrium first, and the size of this neighborhood is measured by a distance function. Without a true distance metric (i.e., without the triangle inequality), the notion of ``neighborhood'' is not well-defined; one cannot say how close is close enough. Thermomajorization gives us an ordering but no neighborhoods. The true distance metric gives us both. A true distance metric guarantees that the direct path between any two distributions is never longer than any detour through a third, producing a stable and consistent geometric space where distances are predictable, shortest paths exist, and the approach to equilibrium can be measured rather than merely ordered.% Since a true distance metric is a valid monotone measure, the thermomajorization framework guarantees a computable lower bound on the thermomajorization crossover time without solving for the full Lorenz curve.
%\iitmd{[MD: Comment: The last sentence may be removed or placed elsewhere as it appears not to show continuation with the previous sentence.]}

\begin{comment}
    Due to trinagular inequality, true distance metric gives  a geometric interpretation that thermomajorization, as a partial order, cannot provide. The thrue distance distance metric from equilibrium traces a geodesic in the Hilbert space and the Mpemba effect corresponds to the hot trajectory's geodesic distance falling below the warm trajectory's geodesic distance at the crossover time. This geometric picture — two systems racing toward equilibrium along geodesics, with the hotter one arriving closer first — is made rigorous and quantitative by the metric structure of Hilbert space in a way that the ordinal language of thermomajorization does not support. In this sense, true distance metric translates the combinatorial statement of thermomajorization into the language of Riemannian geometry, opening a path toward understanding the Mpemba effect through the curvature and geodesic structure of the space of probability distributions.
\end{comment}

%%%%%%%%%%%%%%
\section{Results and Discussion}\label{sec:result}
We are interested in upper and lower bounds on \(r(t) = \DT{h} - \DT{w} \), \(r(t)\) is the difference of distance from equilibrium of a hot system having temperature \(T_h\) and a warm system having temperature \(T_w\) with \(T_h > T_w\). The upper and lower bound on \(r(t)\) is calculated in the Appendix:~\ref{app:bound} and given by equation (\ref{boundcal}) as follows
\[\left|\DT{h}-\DT{w}\right| \leq e^{-Ct} \norm{\rho_h(x,0)-\rho_w(x,0)}_{\omega} .\]
The above equation can be written as
\begin{comment}
\begin{align}\label{ulbound}
    r(t) \geq -e^{-Ct} \left(\DTo{h}+\DTo{w} \right) \\
    r(t) \leq e^{-Ct} \left(\DTo{h}+\DTo{w} \right) 
\end{align}
\end{comment}
\begin{align}
    r(t) &\geq -e^{-Ct} \left(\norm{\rho_h(x,0)-\rho_w(x,0)}_{\omega}  \right) \label{ulbound1} \\
    r(t) &\leq e^{-Ct} \left(\norm{\rho_h(x,0)-\rho_w(x,0)}_{\omega}  \right) \label{ulbound2}.
\end{align}
%If \(r(t)=0\) for some, \(t=t^*\) and some combination of system temperatures \(T_h\) and \(T_w\) then we say that the Mpemba effect is present in the system; otherwise, there is no Mpemba effect.

We say that the Mpemba effect is present if $r(t)$ changes sign at some finite
time $t^{*} \in (0,\infty)$, that is,
\begin{equation}\label{eq:conc_criterion}
  r(t^{*}) = 0, \qquad
  r(t^{*}-\epsilon)\, r(t^{*}+\epsilon) < 0
  \quad 
\end{equation}
for all sufficiently small  \(\epsilon > 0\). We also call this criterion 'inversion' or \(r(t)\) has a root. The requirement $t^{*} < \infty$ is essential, since inequalities~\eqref{ulbound1} and~\eqref{ulbound2} imply, $r(t) \to 0$ as $t \to \infty$ for
every pair of temperatures. Since \(r(0) > 0\), a change of
sign of \(r(t)\) from positive to negative signals the Mpemba effect: the
initially hotter system becomes closer to equilibrium than the initially
warmer one.
%Moreover, $r(t)$ is continuous in $t$, and the norm is Lipschitz. By the intermediate value theorem, a single observation $r(t_1) < 0$ at any time $t_1 > 0$ therefore suffices to establish a crossing in $(0, t_1)$.

\begin{figure}%[t]
  \centering
  \includegraphics[width=\columnwidth]{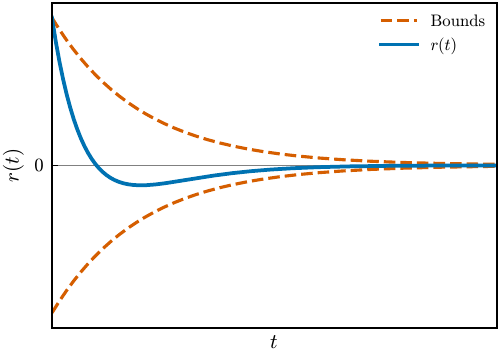}
  \caption{Schematic of the relaxation-difference function $r(t)$ (solid line)
  confined within the exponential envelope $\pm e^{-Ct} \left(\norm{\rho_h(x,0)-\rho_w(x,0)}_{\omega}  \right)$ (dashed lines). Because $|r(t)|$ is bounded and decays to zero,
  any sign change must occur at finite time. The single zero of $r(t)$ marks the
  crossover time $t^{*}$ at which the initially hotter system becomes closer to
  equilibrium than the initially cooler one, which is
  the signature of the Mpemba effect. Axes are in arbitrary units.}
  \label{fig:ulbound}
\end{figure}

The possibility of a root of \(r(t)\) is given schematically in figure~\ref{fig:ulbound}. This dictates the possibility of the presence of the Mpemba effect by the mathematical formulation developed in this present work.

By inequalities~\eqref{ulbound1} and~\eqref{ulbound2}, we can also get crucial dependence on temperatures and potential for occurrence of the Mpemba effect. As $T_w \to T_h$, $\norm{\rho_h(x,0)-\rho_w(x,0)}_{\omega} \to 0$ and hence $r(t) \to 0$ : a Mpemba effect between nearby temperatures is necessarily weak. The potential enters in inequalities~\eqref{ulbound1} and~\eqref{ulbound2} only through
$C$ whose optimal value is spectral gap \cite{LelievreNierPavliotis2013}: strongly confining landscapes with a large spectral gap
suppress any inversion rapidly, whereas landscapes with a small gap, such as shallow or metastable potentials and double wells separated by high barriers, permit long-lived inversions and are therefore the natural candidates for an observable Mpemba effect.

\begin{figure}%[t]
  \centering
  \includegraphics[width=\columnwidth]{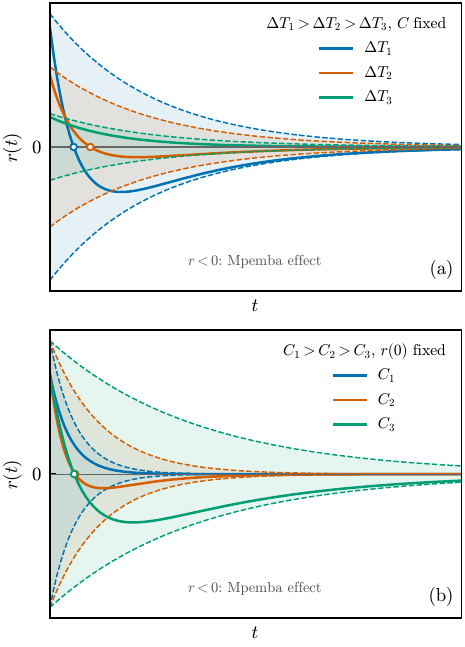}
  \caption{Pictorial representation of the bounds~\eqref{ulbound1}
  and~\eqref{ulbound2} (arbitrary units). Shaded regions bounded by dashed
  lines show \(\pm e^{-Ct}\norm{\rho_h(x,0) - \rho_w(x,0)}_{\omega}\); solid
  lines are schematic trajectories \(r(t)\), and open circles mark the change
  of sign of \(r(t)\) that signals the Mpemba effect. (a)~Fixed \(C\) and
  decreasing temperature difference
  \(\Delta T_1 > \Delta T_2 > \Delta T_3\), where \(\Delta T = T_h - T_w\): the
  Mpemba effect is pronounced for \(\Delta T_1\), weak for \(\Delta T_2\), and
  absent for \(\Delta T_3\). (b)~Fixed \(T_h\), \(T_w\), and hence fixed
  \(r(0)\), with decreasing \(C_1 > C_2 > C_3\): no Mpemba effect for \(C_1\),
  and an increasingly long-lived inversion for \(C_2\) and \(C_3\).}
  \label{fig:mpemba-schematic}
\end{figure}

%%%%%%%%%%%%%%5
Figure~\ref{fig:mpemba-schematic} gives a pictorial representation of these
two dependences. In each panel, the shaded region bounded by dashed lines is
the band
\(|r(t)| \le e^{-Ct}\,\norm{\rho_h(x,0) - \rho_w(x,0)}_{\omega}\) permitted by
the inequalities~\eqref{ulbound1} and~\eqref{ulbound2}, and the solid curve is
a schematic trajectory \(r(t)\) inside it. 
%Since \(r(0) > 0\), a change of sign of \(r(t)\) from positive to negative signals the Mpemba effect: the initially hotter system becomes closer to equilibrium than the initially warmer one. 
%\pcmd{[This sentence can be moved to two paragraphs back while describing Fig.~\ref{fig:ulbound} as inversion is first illustrated there.]} 
Figure~\ref{fig:mpemba-schematic}(a) shows the dependence on the 
temperature difference \(\Delta T = T_h - T_w\) at fixed \(C\). For a large
temperature difference \(\Delta T_1\), the band is wide enough to accommodate
a pronounced and clearly resolved inversion. As \(\Delta T\) decreases, the
band narrows in proportion to \(\norm{\rho_h(x,0) - \rho_w(x,0)}_{\omega}\),
%and so does the largest inversion it can accommodate: 
the inversion is weak for \(\Delta T_2\), 
and for the smallest difference \(\Delta T_3\) no inversion is present. 
Figure~\ref{fig:mpemba-schematic}(b) shows the dependence on \(C\)
for a fixed initial separation, and hence a fixed \(r(0)\). For the largest 
value \(C_1\), the band closes before an inversion can develop, and \(r(t)\)
decays to zero without changing sign, so no Mpemba effect occurs. For smaller
values of \(C\), the band closes more slowly, and an inversion appears for both 
$C_{2}$ and $C_{3}$ and persists for longer time, as for \(C_3\) which has 
the smallest value within the chosen set. 

Both the figures~\ref{fig:ulbound} and ~\ref{fig:mpemba-schematic} are schematic: the 
trajectories are not computed for a specific potential, 
and the bounds determine only the region in which an inversion can occur, 
not whether it occurs. 
However, they provide an understanding that 
the mathematical framework can interpret the possibility of occurrence of the Mpemba 
effect in varied types of systems, in general, as no explicit potential is 
considered in the formulation. Importantly, the schematics depict clearly the 
qualitative effects of the initial temperature differences between the hot and warm 
bath and the potential forms on the chance of observing the Mpemba effect. 
It becomes evident that whether the Mpemba effect can be detected depends on these two 
primary factors: initial temperature difference and the potential structure. 
Therefore, we assert at this point that our present work cannot offer 
a definite prediction of the conditions of emergence of the Mpemba effect.  
However, this general mathematical framework can provide a precise idea about 
the fundamental controlling factors and their qualitative role in the 
possibility of the effect to appear in a given system under some specific conditions. 
%The inferences support the real observations. \pcmd{[To verify whether the above observations have been reported experimentally for some systems, also if theoretical studies exist for some specific systems. In that case, to include references; otherwise, to remove the last line.]}

%%%%%%%%%%%%%%%5

%%%%%%%%%%%%%%%
For the harmonic potential \(V(x) = x^2/2\), the Fokker--Planck operator is
diagonalised exactly by Hermite functions, with eigenvalues \(\lambda_n = -n\),
and every coefficient \(a_n\) in the spectral solution~\eqref{solution} is
available in closed form for Gibbs initial distributions
(Appendix~\ref{app:ou}). For two such initial conditions at temperatures
\(T_h > T_w > 1\), the difference \(r(t) = \mathcal{D}_h(t) - \mathcal{D}_w(t)\) is then given exactly by Eq.~\eqref{eq:ou-r}
\begin{equation*}
  \begin{split}
  r(t) &= \mathcal{D}_h(t) - \mathcal{D}_w(t) \\
       &= \frac{1}{\sqrt{1 - (T_h - 1)^2 e^{-4t}}}
        - \frac{1}{\sqrt{1 - (T_w - 1)^2 e^{-4t}}} .
  \end{split}
\end{equation*}
and series expansion is given by Eq.~\eqref{eq:ou-r-series} 
\begin{equation*}
  \begin{split}
  r(t) = \sum_{k=1}^{\infty} \frac{1}{4^k}\binom{2k}{k}
  &\Bigl[ (T_h - 1)^{2k} - (T_w - 1)^{2k} \Bigr] \\
  &\times e^{-4kt} ,
  \end{split}
\end{equation*}
and also the first few terms are given by Eq.~\eqref{eq:ou-r-terms}
\begin{equation*}
  \begin{split}
  r(t) &= \tfrac{1}{2}\bigl[ (T_h-1)^2 - (T_w-1)^2 \bigr] e^{-4t} \\
  &\quad + \tfrac{3}{8}\bigl[ (T_h-1)^4 - (T_w-1)^4 \bigr] e^{-8t} \\
  &\quad + \tfrac{5}{16}\bigl[ (T_h-1)^6 - (T_w-1)^6 \bigr] e^{-12t}
  + \cdots .
  \end{split}
\end{equation*}
Since $T_h - 1 > T_w - 1 > 0$, every bracket in
Eq.~\eqref{eq:ou-r-series} is strictly positive, and every term is positive
and strictly decreasing in $t$. Hence, for all $t$ at which
$\mathcal{D}_h(t)$ is finite,
\begin{equation*}
  r(t) > 0 \qquad \text{and} \qquad \frac{\dd r}{\dd t} < 0 .
\end{equation*}
The hotter initial state therefore remains farther from equilibrium at all
times, and no Mpemba effect occurs (Appendix.~\ref{app:ou}). No Mpemba effect in case of harmonic potential also reported in \cite{shapira2026}.
%\pcmd{[Is there any previous work that reports that the Mpemba effect is not observed in harmonic potential? In that case, the reference can be mentioned as support for this result. As I find it, the conventional Mpemba effect is not observed for an overdamped Brownian particle in a simple harmonic potential when the initial state is a thermal equilibrium distribution, and the system subsequently evolves under overdamped Langevin dynamics in a common thermal bath. To verify this and to mention that the current mathematical framework can reproduce this finding.]}

%Every term of its mode expansion~\eqref{eq:ou-r-series} is strictly positive, so \(r(t) > 0\) for all \(t\), and the harmonic potential exhibits no Mpemba effect (Sec.~\ref{app:ou}).
%and Fig.~\ref{fig:ou-r}). 
For other confining potentials, we do not have quantitative results, because the coefficients \(a_n\) are not, in general, available
in closed form. They must instead be obtained numerically, for instance by
pseudospectral methods~\cite{bernard, Shizgal2015, Shizgal2016} or by extended
dynamic mode decomposition (EDMD)~\cite{edmd, Klus_2024, ghosh}. Each approach,
however, has a limitation for the present purpose. The pseudospectral scheme
of Ref.~\cite{Shizgal2016} and analytical method \cite{Cukier1973} are formulated for a Dirac delta initial condition
rather than for an arbitrary Gibbs initial distribution, and EDMD, as
implemented in Ref.~\cite{edmd}, yields only data-driven approximations to
the coefficients \(a_n\), not their exact values.

\begin{comment}
    \pcmd{[MD: Section IV can be named as ``Results and Discussion''.\\ 
Comments on section IV.\\
1. According to the present form, the Results section seems very brief. It is suggested to discuss Appendix A here. For Eq.(A2), the final equation can be mentioned in the main text, and the details can be kept as part of the Appendix.\\
2. It is expected that the Results section will contain quantitative data. It is suggested to make Fig.2 quantitative. The results can be presented for several pairs of $T_{h}$ and $T_{w}$ with actual values.\\
3. Discussions on the quantitative calculations of the crossover time can be added, and the results in graphical form can be presented.\\
4. The degree of generality of the results can be discussed.]}
\end{comment}

%%%%%%%%%%%%%%

%%%%%%%%%%%%

%%%%%%%%%%%%%%%%%%
\section{Conclusion}\label{sec:conclusion}

In this work, we addressed a question that had remained open in the theoretical study of the Mpemba effect for Markovian systems under external potentials: whether the very possibility of the phenomenon can be established by mathematical analysis. 
By working in weighted Hilbert spaces, we exploited the fact that the relevant operators become self-adjoint with respect to the weighted complete inner product spaces (i.e., Hilbert spaces), yielding a real and orthogonal spectral decomposition. 
This framework naturally induces a genuine distance metric through the Hilbert space norm, rather than relying on externally introduced pseudo-distance measures.

Using the geometric structure of the weighted Hilbert space together with the Poincar\'e inequality satisfied by the Gibbs equilibrium measure, we established the mathematical possibility for the existence of a crossover in the relaxation dynamics of initially hot and warm states (i.e., the possibility of the Mpemba effect for the overdamped Langevin class of Markovian systems).

Our results complement the thermomajorization approach. While thermomajorization provides an upper bound on the crossover time that is valid for all monotone distance measures, including both true and pseudo-distance measures, it does not by itself establish whether a crossover can actually occur. The present work fills this gap by providing a rigorous functional-analytic criterion for the existence of the phenomenon. We also discuss how the formalism extends to systems with double-well potentials, where the weight function proportional to $\exp[V(x)]$ naturally captures the slow inter-basin relaxation.

Beyond its implications for the Mpemba effect, the weighted Hilbert space framework developed here connects naturally with modern spectral methods for stochastic dynamics. In particular, it provides a mathematically consistent setting for spectral numerical techniques, such as Extended Dynamic Mode Decomposition (EDMD) \cite{edmd, Klus_2024}, pseudo-spectral methods \cite{bernard}, deep learning methods \cite{dlfp}, and dimensionality-reduction methods for the Fokker-Planck equation based on spectral decompositions, such as diffusion maps \cite{diffusionmap}.
%and some other methods given in references \cite{Hongler1982, Caroli1979}.
These connections suggest several promising directions for future research, including obtaining sharper estimates of the crossover time directly from spectral data and extending the present analysis to broader classes of nonequilibrium stochastic systems.
%%%%%%%%%%%%#####################
%%%%%%%%%%%%%%%#######

\begin{acknowledgments}
U. P. acknowledges IIT Mandi for a fellowship. 
M.D. thanks ANRF-SERB (Project No. SRG/2022/000296), 
Department of Science and Technology, Government of India, 
and IIT Mandi (Seed Grant No. IITM/SG/MUD/91) for financial support. 
U. P. acknowledges Abhishek Kaintura, Apurba Biswas, and Sourav Dutta for discussions. Both the authors thank Dr. Harsh Soni for insightful discussions and a thorough reading of the manuscript. 
\end{acknowledgments}
%%%%%%%%%%%%%%

\section*{Declaration}
\textbf{Conflict of interest:} The authors declare that there is no conflict of interest. 
%%%%%%%%%%%%
\appendix
%%%%%%%%%%%%
\section{\label{app:bound} Bounds on \(r(t)\)}
We say the measure \(\mu\) of inner product space \(L^2_{\mu}\) satisfies Poincar\'e inequality with a constant \(C>0\) if, for given any \(h \in L^2(\mu)\) with \(\int h\, \mu dx =0\) following inequality holds \cite{wasser}
\begin{equation}\label{poincare}
    \int h^2\, \mu dx \leq \frac{1}{C} \int \left|\nabla h\right|^2 \, \mu dx.
\end{equation}

Let \(\rho_h(x,0)\), \(\rho_w(x,0)\) are initial conditions for equation (\ref{FPE}) for temperature \(T_h\) and \(T_w\), and corresponding solutions be \(\rho_h(x,t)\), \(\rho_w(x,t)\) respectively. Since \(\rho_{eq}(x)\) it follows the Poincar\'e inequality for unbounded domain \cite[Appendix~A.19]{villani}, \cite{LelievreNierPavliotis2013} and bounded domains  \cite[Section~5.8]{Evans2010}. Considering \(f \in \lpi\) such that \(f = \frac{\rho_h(x,t)-\rho_w(x,t)}{\req} - \int_{\Omega} \left(\frac{\rho_h(x,t)-\rho_w(x,t)}{\req}\right) \, \req dx\), and \(\int_{\Omega} f\, \req dx = 0\), then by using inequality (\ref{poincare}) we get

\begin{comment}
\begin{widetext}
\begin{align}\label{boundcal}
    \int_{\Omega} \left(\rho_h(x,t)-\rho_w(x,t)\right)^2  \, \frac{1}{\req} dx & \leq e^{-2Ct} \quad \int_{\Omega} \left(\rho_h(x,0)-\rho_w(x,0)\right)^2 \, \frac{1}{\req} dx \nonumber \\
    \norm{\left(\rho_h(x,t)-\rho_w(x,t)\right)}_{\omega}^2 &\leq e^{-2Ct} \,\norm{\left(\rho_h(x,0)-\rho_w(x,0)\right)}_{\omega}^2 \nonumber \\
    \norm{\rho_h(x,t)-\rho_w(x,t)}_{\omega} &\leq e^{-Ct} \norm{\rho_h(x,0)-\rho_w(x,0)}_{\omega} \nonumber \\
    \norm{(\rho_h(x,t)-\req)-(\rho_w(x,t)-\req)}_{\omega} &\leq e^{-Ct}  \norm{(\rho_h(x,0)-\req)-(\rho_w(x,0)-\req)}_{\omega} \nonumber \\
    \left|\norm{(\rho_h(x,t)-\req)}_{\omega}-\norm{(\rho_w(x,t)-\req)}_{\omega}\right| &\leq e^{-Ct}  \left(\norm{(\rho_h(x,0)-\req)}_{\omega}+\norm{(\rho_w(x,0)-\req)}_{\omega} \right) \nonumber \\
   \left|\DT{h}-\DT{w}\right| &\leq e^{-Ct} \left(\DTo{h}+\DTo{w} \right)
\end{align}
\end{widetext}
\end{comment}
\begin{comment}
\begin{widetext}
\begin{align}
    -e^{-Ct} \left(\DTo{h}+\DTo{w} \right) \leq (\DT{h}-\DT{w}) \leq e^{-Ct} \left(\DTo{h}+\DTo{w} \right) 
\end{align}
\end{widetext}
\end{comment}

\begin{widetext}
\begin{align}
    \int_{\Omega} \left(\rho_h(x,t)-\rho_w(x,t)\right)^2  \, \frac{1}{\req} dx & \leq e^{-2Ct} \quad \int_{\Omega} \left(\rho_h(x,0)-\rho_w(x,0)\right)^2 \, \frac{1}{\req} dx \nonumber \\
    \norm{\left(\rho_h(x,t)-\rho_w(x,t)\right)}_{\omega}^2 &\leq e^{-2Ct} \,\norm{\left(\rho_h(x,0)-\rho_w(x,0)\right)}_{\omega}^2 \nonumber \\
    \norm{\rho_h(x,t)-\rho_w(x,t)}_{\omega} &\leq e^{-Ct} \norm{\rho_h(x,0)-\rho_w(x,0)}_{\omega} \nonumber \\
    \norm{(\rho_h(x,t)-\req)-(\rho_w(x,t)-\req)}_{\omega} &\leq e^{-Ct}  \norm{\rho_h(x,0)-\rho_w(x,0)}_{\omega} \nonumber \\
    \left|\norm{(\rho_h(x,t)-\req)}_{\omega}-\norm{(\rho_w(x,t)-\req)}_{\omega}\right| &\leq e^{-Ct}  \norm{\rho_h(x,0)-\rho_w(x,0)}_{\omega} \nonumber \\
   \left|\DT{h}-\DT{w}\right| &\leq e^{-Ct} \norm{\rho_h(x,0)-\rho_w(x,0)}_{\omega} \label{boundcal}
\end{align}
\end{widetext}

In above calculations, % we have used two inequalities \(\left| \norm{A} - \norm{B} \right| \leq \norm{A-B}\) and \(\norm{A-B} \leq \norm{A} + \norm{B}\). 
 we have used inequality \(\left| \norm{A} - \norm{B} \right| \leq \norm{A-B}\).
And here \(\DT{h}\) and \(\DT{w}\) are distances from equilibrium for systems with initial temperatures \(T_h\) and, \(T_w\) respectively i.e. \(\DT{h} = d(\rho_{h}(x,t)), \req\) and \(\DT{w} = d(\rho_{w}(x,t), \req\)). 
%%%%%%%%%%%%%
\section{\label{app:oodist} Properties of distance metric for continuous state space system}
%\textbf{Property-\ref{prop:monotone} of section-\ref{sec:metric} : 
\subsection{\(\D\) is monotonically non-increasing with time \(t\)}
Considering \(f \in \lpi\) such that
\begin{align}\label{intfzero}
    f &= \frac{\rho(x,t)-\req}{\req} \nonumber \\
    & \quad - \int_{\Omega} \left(\frac{\rho(x,t)-\req}{\req}\right) \, \req dx,
\end{align}
 and \(\int_{\Omega} f\, \req dx = 0\). 
 %Since \(\rho_{eq}(x)\) it follows the Poincar\'e inequality for unbounded domain \cite[Appendix~A.19]{villani} and bounded domains with reflecting boundaries \cite[Section~5.8]{Evans2010},
 Then by inequality (\ref{poincare})

\begin{equation*}
    \int f^2\, \req dx \leq \frac{1}{C} \int \left|\nabla f\right|^2 \, \req dx.
\end{equation*}
The above inequality leads to 
\begin{equation}\label{deriG}
    \frac{\partial G}{\partial t} \leq -2CG(t)
\end{equation} where 
\begin{align}\label{GD}
    G(t) &= \int_{\Omega} (\rho(x,t)-\req)^2 \, \frac{1}{\req} dx \nonumber \\
    &= (\D)^2 \nonumber \\
    &= (d(\rxt,\req))^2.
\end{align}
The inequality (\ref{deriG}) implies, \(G(t) \leq e^{-2Ct} G(0)\) and this inequality with equation (\ref{GD}) leads to %\(\D &\leq e^{-Ct} \, \mathcal{D}(0)\)
\begin{align}\label{Dbound}
    \D &\leq e^{-Ct}\, \mathcal{D}(0) \nonumber \\
    &  \text{or} \nonumber \\
     d\left(\rxt,\req \right) &\leq e^{-Ct}\, d\left(\rho(x,0),\req \right).
\end{align}
The above inequality, giving an upper bound on the distance from equilibrium.
Considering inequality (\ref{deriG}) %of section: \ref{bound}
\[ \frac{\partial G}{\partial t} \leq -2CG(t)\] where \(G(t) = (\D)^2\). This implies 
\begin{align}
    \frac{\partial (\D)^2}{\partial t} &\leq -2C (\D)^2 \nonumber \\
    2\D \frac{\partial \D}{\partial t} &\leq -2C (\D)^2 \nonumber\\
     \frac{\partial \D}{\partial t} &\leq -C \D  \leq 0,%\\
\end{align}
 so \(\D\) is a non-increasing function.

%\textbf{Property-\ref{prop:bigger} of section-\ref{sec:metric} : If \(T_h > T_w\) then \(\DTo{h} > \DTo{w}\), where \(\DTo{h} \) is distance from equilibrium of system with initial temperature \(T_h\), likewise \(\DTo{w}\)}\\
\subsection{\(\D\) monotonically increases with temperature}
Since \[(\DTo{T})^2= \int_{\Omega}(\rho_T(x,0)-\req)^2\, \frac{1}{\req} dx\]
where \(\rho_T(x,0) = Z_T^{-1} \exp{\frac{-V(x)}{T}}\), \(\req = Z^{-1} \exp{-V(x)}\) and \(Z_T^{-1}\, ,\, Z^{-1}\) are constants such that \(\int_{\Omega} \rho_T(x,0) \, dx = 1\) for \(T > 1\) and \(\int_{\Omega} \req \, dx = 1.\)
We can write \((\DTo{T})^2\) as follows,
\[(\DTo{T})^2 = \int_{\Omega} (f_T-1)^2 \, \req dx \] where \[f_T =\frac{\rho_T(x,0)}{\req} = \frac{Z}{Z_T} \exp{V(x)(1-1/T)}\]
\begin{align*}
    (\DTo{T})^2 &= \int_{\Omega} (f_T^2+1-2f_T) \, \req dx \\
            &= \int_{\Omega} f_T^2\, \req dx  -1 
\end{align*}
\begin{align}\label{eq:deriD}
    \DTo{T} \; \frac{d}{dT} \DTo{T} \;&= \;  \int_{\Omega} \frac{\rho_T}{\req} \frac{\partial \rho_T}{\partial T} \,dx \nonumber\\
        &= \;  \int_{\Omega} \frac{\rho_T}{\req} \frac{\rho_T}{T^2}(V-\langle V \rangle_{\rho_{T}} )\,dx \nonumber\\
         &= \; \frac{1}{T^2} \int_{\Omega} V \frac{\rho_T^2}{\req} \, dx \nonumber\\
         & \quad - \frac{1}{T^2} \langle V \rangle_{\rho_{T}}  \int_{\Omega} \frac{\rho_T^2}{\req} \, dx \nonumber\\
         &= \; \frac{1}{T^2} \int_{\Omega} V f_T \, \rho_T dx \nonumber\\
         & \quad - \frac{1}{T^2} \langle V \rangle_{\rho_{T}}  \int_{\Omega} f_T \, \rho_T dx  \nonumber \\
         &= \; \frac{1}{T^2} \bigl( \langle Vf_T \rangle_{\rho_{T}}  -\langle V \rangle_{\rho_{T}} \langle f_T \rangle_{\rho_{T}} \bigr) \nonumber \\
        &= \;  \frac{1}{T^2} \operatorname{Cov}_{\rho_T} \! \bigl(V,\,f_T \bigr) 
\end{align}
Now we will work out what will be the sign for \(\operatorname{Cov}_{\rho_T} \! \bigl(V,\,f_T \bigr) \). For this we will use the following lemma:
\begin{lemma}[Covariance Identity]\label{lem:cov}
Let $\mu$ be any probability measure on $\Omega$ with density $\rho_\mu(x)$.
For any two functions $\phi(x)$ and $\psi(x)$ that are square-integrable
under $\mu$,
\begin{align}\label{eq:cov_id}
    \operatorname{Cov}_\mu(\phi,\,\psi)
    \equiv 
    \frac{1}{2}\int \! \! \! \int_{\Omega^2} &
    \bigl(\phi(x) - \phi(y)\bigr) \bigl(\psi(x) - \psi(y)\bigr) \nonumber \\
    & \,\rho_\mu(x)\,\rho_\mu(y)\,dx\,dy.
\end{align}
\end{lemma}
\begin{comment}

\begin{proof}
Expand the product in the integrand:
\begin{align*}
    &\frac{1}{2}\iint
    \bigl[\phi(x)\psi(x) - \phi(x)\psi(y)- \phi(y)\psi(x)
    \\
    & \quad + \; \phi(y)\psi(y)\bigr] \rho_\mu(x)\,\rho_\mu(y)\,dx\,dy \\
    &=\;\frac{1}{2}\Bigl[
        2\int\phi\psi\,d\mu
        \;-\; 2\int\phi\,d\mu\int\psi\,d\mu
    \Bigr] \\
    \;&=\; \mathbb{E}_\mu[\phi\psi] - \mathbb{E}_\mu[\phi]\,\mathbb{E}_\mu[\psi],
\end{align*}
where we used $\int \rho_\mu\,dx = 1$ . %\qed
\end{proof}
\medskip
    
\end{comment}
Applying Lemma~\ref{lem:cov} to $\phi = V$ and $\psi = f_T$ under the measure $\rho$:
\begin{align}\label{eq:cov_dbl}
    \operatorname{Cov}_\rho(V,\, f_T) &= \frac{1}{2}\int_{\Omega}\!\!\int_{\Omega} \bigl(V(x)-V(y)\bigr) \nonumber \\
    & \bigl(f_T(x)-f_T(y)\bigr) \,\rho_T(x)\,\rho_T(y)\,dx\,dy.
\end{align}
Since
\begin{equation*}
    f_T(x) \;=\; \frac{Z}{Z_T}
    \exp\!\left(V(x)\Bigl(1-\tfrac{1}{T}\Bigr)\right),
\end{equation*}
which is a strictly increasing function of $V(x)$whenever $T > 1$,
since the exponent $(1-1/T) > 0$.

Therefore, for any pair of points $x, y \in \Omega$:
\begin{align*}
    V(x) > V(y) \;\implies &\; f_T(x) > f_T(y),
    \qquad\text{and}\\
    V(x) < V(y) \;\implies &\; f_T^2(x) < f_T^2(y).
\end{align*}
In both cases the two factors in the integrand of \eqref{eq:cov_dbl}
carry the same sign:
\begin{equation*}
    \bigl(V(x) - V(y)\bigr)\bigl(f_T(x) - f_T(y)\bigr)
    \;\geq\; 0 \qquad \text{for all } x, y \in \Omega,
\end{equation*}
with equality only when $V(x) = V(y)$.
Since $\rho_T(x)\rho_T(y) > 0$ , and  $V$ is non-constant so
%(otherwise $e^{-V}$ would not be normalizable), the double integral in
\eqref{eq:cov_dbl} is strictly positive:
\begin{equation}
    \operatorname{Cov}_\rho\!\bigl(V,\, f_T\bigr) \;>\; 0.
\end{equation}
\medskip
Coming back to equation~(\ref{eq:deriD})
\begin{equation}
    \frac{d}{dT} \DTo{T} \, = \, \frac{1}{T^2 } \frac{1}{\DTo{T}}\operatorname{Cov}_{\rho_T} \! \bigl(V,\,f_T \bigr) > 0
\end{equation}
for all  \(T >1\). Since \(\frac{d}{dT} \DTo{T}\) is strictly positive, for all \(T>1\) so \(\DTo{T}\) is increasing function of \(T\), \(T>1\). This proves property~\ref{prop:bigger} of section~\ref{sec:metric}. 
%%%%%%%%%%%%%%%

%%%%%%%%%%%%%%
% =====================================================================
%  Appendix section: exact relaxation in the harmonic potential
%  Requires: amsmath, amssymb, graphicx; figure file r_t_OU.pdf
%  Macro \fpo (Fokker--Planck operator) as in the main text.
% =====================================================================

\section{Exact relaxation in the harmonic potential}
\label{app:ou}

For the harmonic potential $V(x) = x^2/2$, the Langevin dynamics reduces to
the Ornstein--Uhlenbeck process, and the Fokker--Planck operator
\begin{equation}
  \fpo \rho = \frac{\partial}{\partial x}\bigl( x\, \rho \bigr)
  + \frac{\partial^2 \rho}{\partial x^2}
  \label{eq:ou-fpo}
\end{equation}
can be diagonalised exactly. In this appendix we obtain its eigenvalues and
eigenfunctions, the expansion coefficients of Gibbs initial distributions,
and the distance from equilibrium in closed form, and we show that
no Mpemba effect occurs in this potential.

\subsection{Eigenvalues and eigenfunctions}
\label{app:ou-spectrum}

The stationary distribution of Eq.~\eqref{eq:ou-fpo} is the standard Gaussian,
\begin{equation}
  \req(x) = \frac{1}{\sqrt{2\pi}}\, e^{-x^2/2} .
  \label{eq:ou-stationary}
\end{equation}
On the weighted Hilbert space $L^2_{\omega}(\mathbb{R})$, with
$\omega = 1/\req$ and inner product
\begin{equation}
  \langle f, g \rangle_{\omega}
  = \int_{-\infty}^{\infty} \frac{f(x)\, g(x)}{\req(x)}\, \dd x ,
  \label{eq:ou-inner}
\end{equation}
the operator $\fpo$ is self-adjoint and negative semi-definite, so its
eigenvalues are real and non-positive. We seek solutions of
\begin{equation}
  \fpo \psi_n = \lambda_n\, \psi_n , \qquad \lambda_n \le 0 .
  \label{eq:ou-eigen}
\end{equation}

Writing $\psi = \req f$ and using $\req' = -x\req$, one finds
$(\partial_x + x)(\req f) = \req f'$ and
$\partial_x(\req f') = \req\,(f'' - x f')$. Since
$\fpo = \partial_x(\partial_x + x)$, it follows that
\begin{equation}
  \fpo(\req f) = \req\, \bigl( f'' - x f' \bigr),
  \label{eq:ou-backward}
\end{equation}
and Eq.~\eqref{eq:ou-eigen} becomes $f'' - x f' = \lambda f$. In the
rescaled variable $y = x/\sqrt{2}$ this is
\begin{equation}
  \frac{\dd^2 f}{\dd y^2} - 2y\, \frac{\dd f}{\dd y} - 2\lambda f = 0 ,
  \label{eq:ou-hermite-eq}
\end{equation}
which is the Hermite differential equation. Its solutions with
$\req f \in L^2_{\omega}(\mathbb{R})$ exist only for $-\lambda = n \in \{0,
1, 2, \dots\}$ and are the Hermite polynomials
\begin{equation}
  H_n(y) = (-1)^n\, e^{y^2}\, \frac{\dd^n}{\dd y^n}\, e^{-y^2},
  \label{eq:ou-rodrigues}
\end{equation}
the first few of which are
\begin{equation}
  \begin{aligned}
  H_0 &= 1, & H_1 &= 2y, \\
  H_2 &= 4y^2 - 2, & H_3 &= 8y^3 - 12y, \quad \dots
  \end{aligned}
  \label{eq:ou-hermite-list}
\end{equation}
They satisfy the orthogonality relation
\begin{equation}
  \int_{-\infty}^{\infty} H_m(y)\, H_n(y)\, e^{-y^2}\, \dd y
  = \sqrt{\pi}\, 2^n\, n!\, \delta_{mn} .
  \label{eq:ou-hermite-orth}
\end{equation}

The eigenvalues and normalised eigenfunctions of $\fpo$ are therefore
\begin{equation}
  \begin{split}
  \lambda_n &= -n, \qquad n = 0, 1, 2, \dots, \\
  \psi_n(x) &= \frac{1}{\sqrt{2^n\, n!}}\,
  H_n\!\Bigl(\frac{x}{\sqrt{2}}\Bigr)\, \req(x)
  = \frac{(-1)^n}{\sqrt{n!}}\, \frac{\dd^n \req}{\dd x^n} ,
  \end{split}
  \label{eq:ou-eigenfunctions}
\end{equation}
where the second form follows from Eq.~\eqref{eq:ou-rodrigues}. Using
Eq.~\eqref{eq:ou-hermite-orth} with $y = x/\sqrt{2}$,
\begin{equation}
  \begin{split}
  \langle \psi_m, \psi_n \rangle_{\omega}
  &= \frac{1}{\sqrt{2^{m+n}\, m!\, n!}} \\
  &\quad \times \int_{-\infty}^{\infty} H_m\!\Bigl(\frac{x}{\sqrt{2}}\Bigr)
  H_n\!\Bigl(\frac{x}{\sqrt{2}}\Bigr) \req(x)\, \dd x
  = \delta_{mn},
  \end{split}
  \label{eq:ou-orthonormal}
\end{equation}
so $\{\psi_n\}_{n \ge 0}$ is an orthonormal basis of
$L^2_{\omega}(\mathbb{R})$. The mode $\psi_0 = \req$, with $\lambda_0 = 0$,
is the stationary state, and the spectral gap is $|\lambda_1| = 1$.

Any initial density $\rho(x,0) \in L^2_{\omega}(\mathbb{R})$ then evolves as
\begin{equation}
  \rho(x,t) = \sum_{n=0}^{\infty} a_n\, e^{\lambda_n t}\, \psi_n(x)
  = \sum_{n=0}^{\infty} a_n\, e^{-n t}\, \psi_n(x),
  \label{eq:ou-expansion}
\end{equation}
with expansion coefficients
\begin{equation}
  a_n = \langle \psi_n, \rho(x,0) \rangle_{\omega}
  = \frac{1}{\sqrt{2^n\, n!}}\,
  \Bigl\langle H_n\!\Bigl(\frac{x}{\sqrt{2}}\Bigr) \Bigr\rangle_0 ,
  \label{eq:ou-an-def}
\end{equation}
where $\langle \cdot \rangle_0$ denotes an average over $\rho(x,0)$. In
particular, $a_0 = 1$ by normalisation.

\subsection{Initial conditions and expansion coefficients}
\label{app:ou-coefficients}

We prepare the system in one of two Gibbs states of the harmonic potential at
temperatures $T_h > T_w > 1$,
\begin{equation}
  \rho_i(x,0) = \frac{1}{\sqrt{2\pi T_i}}\, e^{-x^2/(2T_i)},
  \qquad i \in \{h, w\} .
  \label{eq:ou-initial}
\end{equation}
For a Gaussian initial condition of variance $T$, the coefficients follow from
the generating function of the Hermite polynomials,
$\sum_{n} H_n(y)\, s^n/n! = e^{2ys - s^2}$. Setting $y = x/\sqrt{2}$ and
averaging over $\rho(x,0)$, with
$\langle e^{\sqrt{2}\, s x} \rangle_0 = e^{T s^2}$, gives
\begin{equation}
  \begin{split}
  \sum_{n=0}^{\infty} \frac{s^n}{n!}
  \Bigl\langle H_n\!\Bigl(\frac{x}{\sqrt{2}}\Bigr) \Bigr\rangle_0
  &= e^{-s^2} \bigl\langle e^{\sqrt{2}\, s x} \bigr\rangle_0 \\
  &= e^{(T-1)s^2} .
  \end{split}
  \label{eq:ou-generating}
\end{equation}
Comparing powers of $s$, the odd averages vanish and
$\langle H_{2k}(x/\sqrt{2}) \rangle_0 = (2k)!\,(T-1)^k/k!$. Substituting into
Eq.~\eqref{eq:ou-an-def},
\begin{equation}
  a_{2k+1} = 0, \qquad
  a_{2k} = \frac{\sqrt{(2k)!}}{2^k\, k!}\, (T - 1)^k ,
  \label{eq:ou-coefficients}
\end{equation}
for $k = 0, 1, 2, \dots$. Explicitly, for $n \le 10$,
\begin{equation}
  \begin{aligned}
  a_0 &= 1, &
  a_2 &= \tfrac{1}{\sqrt{2}}\,(T-1), \\
  a_4 &= \tfrac{\sqrt{6}}{4}\,(T-1)^2, &
  a_6 &= \tfrac{\sqrt{5}}{4}\,(T-1)^3, \\
  a_8 &= \tfrac{\sqrt{70}}{16}\,(T-1)^4, &
  a_{10} &= \tfrac{3\sqrt{7}}{16}\,(T-1)^5,
  \end{aligned}
  \label{eq:ou-coefficients-list}
\end{equation}
with $a_1 = a_3 = a_5 = a_7 = a_9 = 0$. The odd coefficients vanish because
the initial conditions~\eqref{eq:ou-initial} are even in $x$. In particular,
the slowest mode is not excited ($a_1 = 0$), and the relaxation is governed by
the $n = 2$ mode. 
\begin{comment}
Summing Eq.~\eqref{eq:ou-expansion} with these coefficients (Mehler's formula) shows that $\rho(x,t)$ remains Gaussian, with variance
\begin{equation}
  \sigma^2(t) = 1 + (T - 1)\, e^{-2t} .
  \label{eq:ou-variance}
\end{equation}
\end{comment}
\subsection{Distance from equilibrium}
\label{app:ou-chi2}

We measure the distance from equilibrium
\begin{equation}
  \mathcal{D}(t) = \bigl\| \rho(x,t) - \req \bigr\|_{\omega}^{2}
  = \int_{-\infty}^{\infty}
  \frac{\bigl[\rho(x,t) - \req(x)\bigr]^2}{\req(x)}\, \dd x .
  \label{eq:ou-D-def}
\end{equation}
By Parseval's identity and the orthonormality~\eqref{eq:ou-orthonormal}, it
diagonalises in the eigenbasis,
\begin{equation}
  \mathcal{D}(t) = \sum_{n=1}^{\infty} a_n^2\, e^{2\lambda_n t}
  = \sum_{n=1}^{\infty} a_n^2\, e^{-2 n t} .
  \label{eq:ou-parseval}
\end{equation}
Inserting Eq.~\eqref{eq:ou-coefficients}, with
$a_{2k}^2 = 4^{-k}\binom{2k}{k}(T-1)^{2k}$, and setting
$u(t) = (T-1)^2 e^{-4t}$,
\begin{equation}
  \mathcal{D}(t) = \sum_{k=1}^{\infty} \frac{1}{4^k}\binom{2k}{k}\, u^k
  = \frac{1}{\sqrt{1 - (T - 1)^2 e^{-4t}}} - 1 ,
  \label{eq:ou-D-closed}
\end{equation}
where we used the binomial series
$\sum_{k \ge 0} \binom{2k}{k} y^k = (1 - 4y)^{-1/2}$, valid for $|y| < 1/4$.
The distance from equilibrium is therefore finite if and only if $(T - 1)\, e^{-2t} < 1$.
For $1 < T < 2$ this holds for all $t \ge 0$; for $T \ge 2$, $\mathcal{D}(t)$
is infinite for $t \le \tfrac{1}{2}\ln(T - 1)$ and finite thereafter.
\begin{comment}
As an independent check, the Gaussian form~\eqref{eq:ou-variance} gives
directly
\begin{equation}
  \mathcal{D}(t) + 1
  = \int_{-\infty}^{\infty} \frac{\rho(x,t)^2}{\req(x)}\, \dd x
  = \frac{1}{\sqrt{\sigma^2 \bigl( 2 - \sigma^2 \bigr)}} ,
  \label{eq:ou-direct}
\end{equation}
and $\sigma^2(2 - \sigma^2) = 1 - (T-1)^2 e^{-4t}$, in agreement with
Eq.~\eqref{eq:ou-D-closed}.
\end{comment}
\subsection{Absence of the Mpemba effect}
\label{app:ou-mpemba}

Let $\mathcal{D}_h(t)$ and $\mathcal{D}_w(t)$ denote the distance from equilibrium for the
hot and warm initial conditions~\eqref{eq:ou-initial}. The Mpemba effect if and only if a change of sign, at some finite time, of the
difference
\begin{equation}
  \begin{split}
  r(t) &= \mathcal{D}_h(t) - \mathcal{D}_w(t) \\
       &= \frac{1}{\sqrt{1 - (T_h - 1)^2 e^{-4t}}}
        - \frac{1}{\sqrt{1 - (T_w - 1)^2 e^{-4t}}} .
  \end{split}
  \label{eq:ou-r}
\end{equation}
From Eq.~\eqref{eq:ou-D-closed}, its mode expansion is
\begin{equation}
  \begin{split}
  r(t) = \sum_{k=1}^{\infty} \frac{1}{4^k}\binom{2k}{k}
  &\Bigl[ (T_h - 1)^{2k} - (T_w - 1)^{2k} \Bigr] \\
  &\times e^{-4kt} ,
  \end{split}
  \label{eq:ou-r-series}
\end{equation}
whose first terms are
\begin{equation}
  \begin{split}
  r(t) &= \tfrac{1}{2}\bigl[ (T_h-1)^2 - (T_w-1)^2 \bigr] e^{-4t} \\
  &\quad + \tfrac{3}{8}\bigl[ (T_h-1)^4 - (T_w-1)^4 \bigr] e^{-8t} \\
  &\quad + \tfrac{5}{16}\bigl[ (T_h-1)^6 - (T_w-1)^6 \bigr] e^{-12t}
  + \cdots .
  \end{split}
  \label{eq:ou-r-terms}
\end{equation}

Since $T_h - 1 > T_w - 1 > 0$, every bracket in
Eq.~\eqref{eq:ou-r-series} is strictly positive, and every term is positive
and strictly decreasing in $t$. Hence, for all $t$ at which
$\mathcal{D}_h(t)$ is finite,
\begin{equation}
  r(t) > 0 \qquad \text{and} \qquad \frac{\dd r}{\dd t} < 0 .
  \label{eq:ou-r-positive}
\end{equation}
The hotter initial state therefore remains farther from equilibrium at all
times, and no Mpemba effect occurs. At late times $r(t)$ is dominated by the
$k = 1$ term,
\begin{equation}
  r(t) \simeq \tfrac{1}{2}\,(T_h - T_w)(T_h + T_w - 2)\, e^{-4t},
  \qquad t \to \infty ,
  \label{eq:ou-r-asymptotic}
\end{equation}
which decays at the rate $4 = 2|\lambda_2|$, because $a_1 = 0$ and the
distance from equilibrium is quadratic in the amplitudes.
\bibliography{apssamp}% Produces the bibliography via BibTeX.

\end{document}